\documentstyle[epsf,preprint,aps]{revtex}
\draft
\preprint{SNUTP 00/004}
\begin{document}
\title{\Large\bf Various Modified Solutions of the Randall-Sundrum 
Model with the Gauss-Bonnet Interaction}
\author{Jihn E. Kim,\footnote{jekim@phyp.snu.ac.kr} 
Bumseok Kyae and Hyun Min Lee} 
\address{ Department of Physics and Center for Theoretical
Physics, Seoul National University,
Seoul 151-742, Korea}
\maketitle

\begin{abstract} 
The Gauss-Bonnet interaction is the only consistent quadratic
interaction below the Planck scale in the Randall-Sundrum 
compactification. We study various static and inflationary
solutions including this Gauss-Bonnet interaction.
\end{abstract}


\newpage

\section{introduction}

Recently, Randall and Sundrum(RS) proposed a compactification
scheme with nonvanishing cosmological constant in the
bulk~\cite{rs} which has immediately attracted a great deal of
attention~\cite{kal,nihei,kimkim,kkl}. The most simple compactification
studied in superstring models before the RS proposal has been the 
orbifold compactification~\cite{orb} in which the compactified space 
is flat. On the other hand, the Randall-Sundrum compactification 
allows a nonflat compactified space, but the analysis is relatively 
simple. Because of the nonflat nature of the bulk between two branes,
there exists an exponential warp factor for metrics going from
one brane to the other~\cite{rs}. This exponential warp factor
has been suggested for a large hierarchy between the Planck scale
$M_P=2.44\times 10^{18}$~GeV and the electroweak scale $v\simeq$~250~GeV.

Among two branes, let Brane 1 (B1) the hidden-sector brane and
Brane 2 (B2) the visible sector brane. An exponential warp factor
suppresses the soft mass in the visible brane B2, and it is possible to
obtain this small ratio because the Higgs mass term at B2 is a
dimension two operator. Thus, in the RS world, one
changes the traditional gauge hierarchy problem to a problem
in geometry. Using the same argument, the nonrenormalizable
operators are suppressed not by $M_P$ but by $v$. 
Thus one has to make sure that the theory has a high degree of 
symmetry to suppress sufficiently the unwanted operators.

Another problem is the problem of inflation. 
Generally, inflation occurs unless one fine-tunes the bulk 
cosmological constant and the brane tensions~\cite{kal,nihei,kimkim}. 
For the fine-tuned relations~\cite{nihei,kimkim,kkl}, there exist 
static solutions. So far it has not been shown that any of the static
solutions is the $t\rightarrow\infty$ limit, not allowing a 
graceful exit from the
inflationary period. In addition, there is a possibility that the
separation between the branes is expanding or shrinking exponentially.
However, this last problem may be understood by introducing a 
scalar field in the bulk~\cite{wise}. 

The most interesting point of the RS world is the interplay of the
bulk and the brane world. In particular, the bulk cosmological
constant ($k$) and the brane tensions ($k_i\ (i=1,2)$) must be 
related, $k_1=k=-k_2$ and $k_1>0$. But the
expansion rate of the observable universe is measured by the 
Hubble parameter which is a function of $k$ and $k_2$.
These $k$'s are the appropriately
defined from the original bulk cosmological constant $\Lambda_b$ and the
brane tensions $\Lambda_1, \Lambda_2$ at B1 and B2~\cite{inflation}. 
One interesting point of the RS compactification is that there
may exists a possibility of understanding 
the old cosmological costant problem.

Below the Planck scale, the higher order effective interaction in the
RS model is known to be the Gauss-Bonnet interaction~\cite{kkl}.
In contrast to the models without the Gauss-Bonnet interaction,
this model allows solutions with a positive $\Lambda_2$ at the
visible brane, which is suitable for a proper expansion in the
standard big bang cosmology after the inflationary period.

In Sec. II, the RS compactification with the Gauss-Bonnet term
is explored. In Sec. III, the static background solutions with the
Gauss-Bonnet interaction are presented. In Sec. IV, simple inflationary
solutions are given. In Sec. V, we present other possible inflationary 
solutions. In Sec. VI, metric perturbation near the static 
background geometry is discussed.

\section{gauss-bonnet interaction}

We will neglect the matter interaction, and consider only
the gravitational interaction with cosmological constants
in the bulk and at the branes. The space-time dimension is
$D=5$. The fifth dimension $x^4\equiv y$ is compactified with
an $S_1/Z_2$ orbifold.
The five dimensional index is $M, N = 0, 1, \cdots,
4$ and the four dimensional brane world index is $\mu, \nu=0, 1, 
\cdots, 3$. The fifth dimension variable $y$ ranges in the region 
[0, 1/2]. The $S_1/Z_2$ orbifold is used to
locate the two branes at $y=0$ and $y=1/2$. The periodicity 
of $y$ is 1.

Below the Planck scale, the higher order gravity effects can be 
added as effective interaction terms. Since we are neglecting the
matter interactions, the possible terms in the Lagrangian is, up to 
O($R^2/M^2$)
\begin{eqnarray}
\label{action}
S&=&\int d^5x\sqrt{-g}\left( {M^3\over 2}R-\Lambda_b +\frac{1}{2}\alpha M R^2
+\frac{1}{2}\beta M R_{MN}R^{MN}+\frac{1}{2}\gamma MR_{MNPQ}R^{MNPQ}\right)
\nonumber \\
&&+\sum_{i=1,2\ {\rm branes}}\int d^4x\sqrt{-g^{(i)}}\left({\cal L}_i
- \Lambda_i\right)
\end{eqnarray}  
where $g, g^{(i)}$ are the determinants of the metrics in the
bulk and the branes, $M$ is the five dimensional gravitational 
constant, $\Lambda_b$
and $\Lambda_i$ are the bulk and brane cosmological constants,
and $\alpha,\beta,\gamma$ are the effective couplings.
We assume that the three dimensional space is homogeneous and 
isotropic, and hence the metric is parametrized by $n, a$, and $b$
\begin{equation}\label{metric}
ds^2=-n^2(\tau,y)d\tau^2+a^2(\tau,y)\delta_{ij}dx^idx^j+b^2(\tau,y)dy^2.
\end{equation}
where the Roman characters $i,j$ denote the space 
indices 1, 2, and 3.

It has been found that there exist solutions consistent with the 
Randall-Sundrum setup if the additional interaction is of
the Gauss-Bonnet type~\cite{kkl}, namely
$\beta=-4\alpha$ and $\gamma=\alpha$ are satisfied. In this
case, the higher $y$ derivative terms, 
$n^{\prime\prime\prime\prime}$,
$n^{\prime\prime\prime}n^\prime$, $n^{\prime\prime}n^{\prime\prime}$,
$a^{\prime\prime\prime\prime}$,
$a^{\prime\prime\prime}a^\prime$, $a^{\prime\prime}
a^{\prime\prime}\cdots$,\footnote{Here, $\prime$ denotes the 
derivative with respect to $y$.} are absent in the 
left-hand side of the Einstein equation.
These conditions for vanishing higher $y$ derivative terms are  
necessary since the right-hand side of the
Einstein equation contains only
one power of the Dirac delta function and the higher derivative
terms diverge more rapidly than the delta function at the
branes. We find that this result is highly nontrivial.  

However, it can be anticipated from the fact that the Gauss-Bonnet term
can be  rewritten as a pseudoscalar quantity,
$\epsilon^{MNOP}\epsilon_{QSTU}R_{MN}\,^{QS}R_{OP}\,^{TU}$.
>From the antisymmetric property of the Riemann tensor, we observe that 
there are no $(a'')^2$, $(n'')^2$ and $a''n''$ terms
\footnote{Under the metric assumption Eq.~(2), 
the terms including $b''$ are 
absent in the action.  We can see it from the antisymmetric 
property of the Riemann tensor, $R^M \, _{NST}=\partial_S
{\Gamma^M}_{TN}-\cdots$ and unique non-vanishing Christoffel symbol 
containing the first derivative of $b$ 
with respect to $y$ is ${\Gamma^5}_{55}=\frac{b^\prime}{b}$.}
in the action which would have given 
the unwanted fourth order derivatives in the equations of motion. 
Thus, the Gauss-Bonnet effective interaction does not contain
higher $y$ derivatives.

The Gauss-Bonnet term $E=R^2-4R_{MN}R^{MN}+R_{MNPQ}R^{MNPQ}$ is 
a total derivative in $D=4$ spacetime, 
in which case it does not change the Einstein gravity. On the other hand, 
for $D\ne 4$ it is not a topological quantity any more. Still,
it does not contribute to the massive poles of the spin-2 
propagator~\cite{ghost,deser}. It means that the metric variations 
near the flat space do not give rise to a ghost graviton even with 
the Gauss-Bonnet term. In general, a combination of the 
quadratic curvature terms without the Gauss-Bonnet ratio leads to 
ghosts. But it may not be meaningful if the location of the 
ghost pole in the graviton propagator is above the Planck scale 
where the derivative expansion breaks down.   
However, the Gauss-Bonnet term possibly 
excites ghost particles near anti-de Sitter space 
in the sense that the sign of the propagator can be 
flipped~\cite{deser}. 

The general curvature squared 
terms in any space-time dimension $D$ can be 
rewritten as~\cite{shapiro},
\begin{equation} \label{squared}
\alpha R^2+\beta R_{MN}R^{MN}+\gamma R_{MNPQ}R^{MNPQ}
\end{equation} 
\begin{equation}
=-\bigg[\frac{(D-2)\beta+4\gamma}{4(D-3)}\bigg]E
+\bigg(\frac{D-2}{D-3}\bigg)(\beta /4+\gamma)C^2
+\bigg[\frac{4(D-1)\alpha+D\beta+4\gamma}{4(D-1)}\bigg]R^2 \\ 
\end{equation}
where $E$ is the Gauss-Bonnet term, $C^2$ is the square of 
the Weyl tensor as follows, 
\begin{eqnarray}
E&=&R^2-4R_{MN}R^{MN}+R_{MNPQ}R^{MNPQ}, \\
C^2&=&R_{MNPQ}^2-\frac{4}{D-2}R_{MN}^2+\frac{2}{(D-1)(D-2)}R^2.
\end{eqnarray}
Note that if the metric is conformally flat 
and $16\alpha+5\beta+4\gamma=0$ in $D=5$,  
the curvature squared terms appear necessarily 
 in the Gauss-Bonnet combination 
because the Weyl tensor vanishes for a 
conformally flat metric, $n(\tau,y)=a(\tau,y)=b(\tau,y)$.  
Then the coefficient of the resultant Gauss-Bonnet term becomes 
$(8\alpha+\beta)/4$.

For $n(\tau,y)=a(\tau,y)$ 
and $16\alpha+5\beta+4\gamma=0$, there still exist higher time 
derivatives $\sim (4\alpha+\beta)(\ddot{a}/a-\ddot{b}/b)^2$ 
in the action from the curvature squared terms.
Thus, to eliminate higher time derivatives too, 
in addition to the condition $16\alpha+5\beta+4\gamma=0$, 
we should choose the Gauss-Bonnet form in curvature squared terms
or conformally flat metric, $n(\tau,y)=a(\tau,y)=b(\tau,y)$.

Variations of the above action with the Gauss-Bonnet 
term gives, apart from those for the brane Lagrangian,
$$
\sqrt{-g}\bigg[R_{MN}-{1\over 2}g_{MN}R-\frac{\alpha}{2 M^2}g_{MN}\left(
R^2-4R_{PQ}R^{PQ}+R_{STPQ}R^{STPQ}\right)
$$
$$
+\frac{2\alpha}{M^2}\left(RR_{MN}-4R_{MP}R_N\,^P+R_{MQSP}R_N\,^{QSP}\right)
+\frac{2\alpha}{M^2}\left(g_{MN}R_{;P}\,^{;P}-R_{;M;N}\right)
$$
\begin{equation} \label{eom}
-\frac{4\alpha}{M^2}\left(g_{MN}R^{PQ}\,_{;P;Q}+R_{MN;P}\,^{;P}
-R_M\,^P\,_{;N;P}-R_N\,^P\,_{;M;P}\right)
\end{equation}
$$
+\frac{2\alpha}{M^2}\left(R_M\,^P\,_N\, ^Q\,_{;P;Q}+R_M\,^P\,_N\,^Q 
\,_{;Q;P}\right)\bigg]
$$
$$
=-M^{-3}\left[\Lambda_b\sqrt{-g}g_{MN}+\Lambda_1\sqrt{-g^{(1)}}
g^{(1)}_{\mu\nu}\delta^\mu_M\delta^\nu_N\delta(y)+\Lambda_2\sqrt{-g^{(2)}}
g^{(2)}_{\mu\nu}\delta^\mu_M\delta^\nu_N\delta(y-\frac{1}{2})\right]
$$
where 1 refers to the brane of the hidden world B1 and 2 refers to
the visible brane B2. The left-hand side of the above equation 
contains the extra term due to the Gauss-Bonnet term,
$\sqrt{-g}X_{MN}$, in addition to the familiar Einstein 
tensor $\sqrt{-g}G_{MN}$. With the metric given in Eq.~(\ref{metric}), 
the $G_{MN}$ and $X_{MN}$ are
\begin{eqnarray}
\label{Ee:00} 
G_{00}&=&3\frac{\dot a}{a} \left(\frac{\dot
a}{a}+\frac{\dot b}{b}\right)
-\frac{3 n^2}{b^2}
\left[\frac{a''}{a}+\frac{a'}{a}\left(\frac{a'}{a}-\frac{b'}{b}\right)\right]
\\
\label{Ee:ii}
G_{ii}&=&-\frac{a^2}{n^2} \left[2\frac{\ddot a}{a}+\frac{\ddot b}{b}
-\frac{\dot a}{a}\left(2\frac{\dot n}{n}-\frac{\dot a}{a}\right)
-\frac{\dot b}{b}\left(\frac{\dot n}{n}-2\frac{\dot a}{a}\right)\right]
\nonumber\\&&
+\frac{a^2}{b^2} \left[\frac{n''}{n}+2\frac{a''}{a}
+\frac{a'}{a}\left(2\frac{n'}{n}+\frac{a'}{a}\right)
-\frac{b'}{b}\left(\frac{n'}{n}+2\frac{a'}{a}\right)\right]
\\
\label{Ee:55}
G_{55}&=&-\frac{3 b^2}{n^2} \left[\frac{\ddot a}{a}
-\frac{\dot a}{a}\left(\frac{\dot n}{n}-\frac{\dot a}{a}\right)\right]
+3\frac{a'}{a}\left(\frac{n'}{n}+\frac{a'}{a}\right)
\\
\label{Ee:05}
G_{05}&=&3\left(\frac{\dot a}{a}\frac{n'}{n}+\frac{\dot b}{b}\frac{a'}{a}
-\frac{\dot a'}{a}\right)
\end{eqnarray}

\begin{eqnarray}
X_{00}&=&\frac{12\alpha}{M^2}\bigg(\frac{\dot{a}^3\dot{b}}{a^3 n^2 b}
-\frac{\dot{a}^2 a^{\prime\prime}}{a^3 b^2}
+\frac{\dot{a}^2 a^\prime b^\prime}{a^3 b^3}
-\frac{\dot{a}\dot{b}{a^\prime}^2}{a^3 b^3}
+\frac{{a^\prime}^2 a^{\prime\prime}n^2}{a^3 b^4}
-\frac{{a^\prime}^3 b^\prime n^2}{a^3 b^5}\bigg)\\
X_{ii}&=&\frac{4\alpha}{M^2}\bigg(-2\frac{\ddot{a}\dot{a}\dot{b}}{n^4 b}
+3\frac{{\dot{a}}^2\dot{n}\dot{b}}{n^5 b}
+2\frac{\ddot{a}a^{\prime\prime}}{n^2 b^2}
-2\frac{\ddot{a}a^\prime b^\prime}{n^2 b^3}
+\frac{\ddot{b}{a^\prime}^2}{n^2 b^3}
+\frac{\dot{a}^2 n^{\prime\prime}}{n^3 b^2}\\ \nonumber
&&-\frac{\dot{a}^2 b^{\prime\prime}}{n^4 b}
-2\frac{\dot{a}^2 {n^\prime}^2}{n^4 b^2}
-\frac{\dot{a}^2 n^\prime b^\prime}{n^3 b^3}
-2\frac{\dot{a}\dot{n}a^{\prime\prime}}{n^3 b^2}
+2\frac{\dot{a}\dot{n}a^\prime b^\prime}{n^3 b^3}
-2\frac{\dot{b}^2 {a^\prime}^2}{n^2 b^4}\\ \nonumber
&&-2\frac{\dot{a}\dot{b}a^\prime n^\prime}{n^3 b^3}
-\frac{\dot{b}\dot{n}{a^\prime}^2}{n^3 b^3}
-2\frac{(\dot{a}^\prime)^2}{n^2 b^2}
+4\frac{\dot{a}^\prime \dot{a}n^\prime}{n^3 b^2}
+4\frac{\dot{a}^\prime \dot{b}a^\prime}{n^2 b^3}  
-2\frac{a^{\prime\prime}a^\prime n^\prime}{n b^4}\\ \nonumber
&&-\frac{n^{\prime\prime}{a^\prime}^2}{n b^4}
+3\frac{{a^\prime}^2 n^\prime b^\prime}{n b^5}\bigg)\\
X_{55}&=&\frac{12\alpha}{M^2}\bigg(\frac{\dot{a}^3\dot{n}b^2}{a^3 n^5}
-\frac{\ddot{a}\dot{a}^2 b^2}{a^3 n^4}
+\frac{\ddot{a}{a^\prime}^2}{a^3 n^2}
+\frac{\dot{a}^2 a^\prime n^\prime}{a^3 n^3}
-\frac{\dot{a}\dot{n}{a^\prime}^2}{a^3 n^3}
-\frac{{a^\prime}^3 n^\prime}{a^3 n b^2}\bigg)\\
X_{05}&=&\frac{12\alpha}{M^2}\bigg(\frac{\dot{a}^3 n^\prime}{a^3 n^3}
+\frac{\dot{a}^2\dot{b}a^\prime}{a^3 n^2 b}
-\frac{\dot{a}{a^\prime}^2 n^\prime}{a^3 n b^2}
-\frac{\dot{b}{a^\prime}^3}{a^3 b^3}
-\frac{\dot{a}^2\dot{a}^\prime}{a^3 n^2}
+\frac{\dot{a}^\prime {a^\prime}^2}{a^3 b^2}\bigg)
\end{eqnarray}
where $\prime$ denotes the derivative with repect to $y$ and 
$\cdot$ denote the derivatives with respect to $\tau$.

The equation of motions Eq.~(7) are
\begin{equation}\label{Eeqn}
G_{MN}+X_{MN}=T_{MN}
\end{equation} 
where
\begin{eqnarray}
T_{00}&=&\frac{n^2}{M^3}\left[\Lambda_b+\frac{\delta\left(y\right)}{b}\Lambda_1
+\frac{\delta\left(y-\frac{1}{2}\right)}{b}\Lambda_2\right] \nonumber \\
T_{ii}&=&-\frac{a^2}{M^3}\left[\Lambda_b+\frac{\delta\left(y\right)}{b}\Lambda_1
+\frac{\delta\left(y-\frac{1}{2}\right)}{b}\Lambda_2\right] \nonumber \\
T_{55}&=&-\frac{b^2}{M^3}\Lambda_b \nonumber \\
T_{05}&=&0. 
\end{eqnarray} 

\section{static solutions}

To find static solutions, let us assume that the metric
takes the following form,
\begin{equation}
ds^2=e^{-2\sigma(y)}\eta_{\mu\nu}dx^\mu dx^\nu+b_0^2 dy^2
\end{equation}
where the length parameter $b_0$ is a constant. 
Note that the modified Einstein equation for the (00) 
component is identical to the $(ii)$ component in Eq.~(\ref{Eeqn}). 
Thus, the (00), $(ii)$, and (55) components of Eq.~(\ref{Eeqn}) lead to
two equations,
\begin{equation}
{3\sigma^{\prime\prime}\over b_0^2}\left(1-{4\alpha\over M^2b_0^2}
(\sigma^\prime)^2\right)=
{\Lambda_1\over M^3 b_0}\delta(y)+{\Lambda_2\over 
M^3 b_0}\delta(y-\frac{1}{2})
\end{equation}
\begin{equation}\label{sigma}
{6(\sigma^\prime)^2\over b_0^2}\left(
1-{2\alpha\over M^2 b_0^2}(\sigma^\prime)^2\right)={-\Lambda_b\over M^3}
\end{equation} 

There exist two solutions of Eq.~(\ref{sigma}), consistent with the
orbifold symmetry $y\rightarrow -y$,
\begin{equation}
\sigma^{\pm}=b_0|y|\left[{M^2\over 4\alpha}\left(
1\pm \left(1+{4\alpha\Lambda_b\over 3M^5}\right)^{1\over 2}\right)\right]
^{1\over 2}\equiv k_{\pm}b_0|y|.
\end{equation}
Let us call $\sigma^+$(or $k_+$) and $\sigma^-$(or $k_-$) solutions $+$
and $-$ solutions, respectively.
{\it Note that both positive and negative bulk cosmological constants are 
possible for the + solution $\sigma^+$.} 
If $\Lambda_{b}=0$, we have an AdS space for the $+$ 
solution irrespective of $\alpha$,
and a Minkowski space for the -- solution~\cite{deser}.
The RS solution is obtained by taking
$\alpha\rightarrow 0$ in the -- solution.

These solutions exist for:\\  
\indent (i) $\alpha<0$ and $\Lambda_b<0$ allows only $\sigma^-$, and\\
\indent (ii) $\alpha>0$ allows both $\sigma^\pm$ solutions. The $\sigma^+$ 
solution is possible for both $\Lambda_b>0$ and\\
\indent $\Lambda_b<0$.  The
$\sigma^-$ solution is possible only for $\Lambda_b<0$.
In any case, there exists the lower 
\indent limit of $\alpha\Lambda_b$,
$\alpha\Lambda_b\ge -3M^5/4$. 

Comparing our results with that of Randall and Sundrum 
$k=(-\Lambda_b/6M^3)^{1/2}$,
the `effective' bulk cosmological constant by the Gauss-Bonnet interaction
can be defined as 
\begin{equation} \label{blcos}
\Lambda^{(b)}_{eff}
\equiv -\frac{3M^5}{2\alpha}\left(
1\pm\sqrt{1+\frac{4\alpha\Lambda_b}{3M^5}}\right)
=-6M^3k_{\pm}^2.
\end{equation}
This is because our geometry of the AdS space guarantees 
a negative bulk cosmological constant effectively. 
For $\sigma^{\pm}$ to be a 
real number it should have the negative sign, which is
the same as in the RS case.

Considering the discontinuities at the branes,  
\begin{equation}\label{disconti}
|y|'=2\left(\theta(y)-\theta(y-\frac{1}{2})\right)-1,
\end{equation}
which comes from the periodicity in $y$ direction and the orbifold symmetry,
we obtain two solutions if the following relations among the brane
cosmological constants are satisfied
\begin{eqnarray}\label{finetun}
\Lambda_1^{\mp}=-\Lambda_2^{\pm}&=&\mp6k_{\pm}M^3
\sqrt{1+\frac{4\alpha\Lambda_b}{3M^5}}\nonumber \\
&=&\mp 6M^3\left[{M^2\over 4\alpha}\left(1\pm \left(
1+{4\alpha\Lambda_b\over 3M^5}\right)^{1\over 2}\right)
\left(1+{4\alpha\Lambda_b\over 3M^5}\right)\right]^{1\over 2}
\end{eqnarray}
where $k_\pm>0$. 
The RS solution is obtained by taking
$\alpha\rightarrow 0$ in the -- solution.
{\it Note that the visible brane can take a positive 
cosmological constant.}  It will be important in the later stage
of the evolution of the universe,
which will be considered in the following section.

Possible solutions are depicted in Fig.~1 as a function of the
Gauss-Bonnet coupling $\alpha$. The vertical axis$(\equiv \lambda_2)$ 
is the solution for $\Lambda_2$ in the visible brane in units of
$\sqrt{6M^3|\Lambda_b|}$, and the horizontal axis $(\equiv \alpha_\Lambda)
$ is defined as $4\alpha\Lambda_b/(3M^5)$.

Similarly, we can define the `effective' brane cosmological constants as
\begin{equation} \label{brcos}
\Lambda_{eff(i)}^{\pm}\equiv 
\frac{\Lambda_{i}^{\pm}}{\sqrt{1+4\alpha\Lambda_b/3M^5}}
\equiv 6M^3k_{i,\pm}.
\end{equation}
Thus we see the RS-like fine tuning conditions again,
\begin{equation}
\frac{\Lambda_{eff(1)}^{\mp}}{6M^3}=-\frac{\Lambda_{eff(2)}^{\pm}}{6M^3}
=\mp \sqrt{\frac{-\Lambda_{eff}^{(b)}}{6M^3}}.
\end{equation}  

The Planck constant at B2(visible sector) is given by \cite{rs}
\begin{eqnarray} \label{plank}
{M_{P}}^2&=&M^3b_0\int_{-\frac{1}{2}}^{\frac{1}{2}}dy 
e^{-2k_{\pm}b_0|y|}=\frac{M^3}{k_{\pm}}[1-e^{-k_{\pm}b_0}] \nonumber \\
&=&M^2\bigg[\frac{1}{4\alpha}\bigg(1
\pm\Big(1+\frac{4\alpha \Lambda_{b}}{3M^5}\Big)^{\frac{1}{2}}
\bigg)\bigg]^{-\frac{1}{2}}[1-e^{-k_{\pm}b_0}].
\end{eqnarray}
The Higgs boson mass parameter at the visible sector is obtained
by redefining the Higgs field such that the kinetic energy
term of the Higgs boson takes a standard form \cite{rs}. Thus the Higgs
mass parameter is given by
\begin{eqnarray} \label{tev}
m&\equiv& e^{-\frac{b_0}{2}k_{\pm}}m_0\nonumber\\
&=&m_0\, \exp\bigg(-\frac{b_0}{2}\bigg[\frac{M^2}
{4\alpha}\bigg(1\pm\Big(1+\frac{4\alpha 
\Lambda_{b}}{3M^5}\Big)^{\frac{1}{2}}\bigg)\bigg]^{\frac{1}{2}}\bigg)
\end{eqnarray}
where $m_0$ is the mass given in the fundamental Lagrangian, before
redefining the Higgs field.

For $k_+b_0\simeq 74$, the + solution gives a needed large
mass hierarchy through the warp factor $e^{-\frac{b_0}{2}k_+}$
from the input mass parameter($M$)  of order
$10^{19}$~GeV, leading to a TeV scale observable mass. 
To achieve a sufficient hierarchy, Randall and Sundrum 
set $k^2=-\Lambda_b/6M^3\approx M^2\approx M_{P}^2$ and 
$b_0/2\approx 37/M$. 
These results can be reproduced for our $+$ solution with $\alpha=O(1)$ and 
$\Lambda_b=O(1)\times M^5$. For example,
$k_+=M\approx M_P$ for 
$\Lambda_b/M^5=12\alpha-6$ and $\alpha>\frac{1}{4}$.
In this case, the brane cosmological constants are given
by $\Lambda_1^{-}=-\Lambda_2^{+} =-6M^4|4\alpha-1|$.    
Thus, we have to set $b_0/2=37/M$ to explain 
the hierarchy between the Planck and TeV scale.  
Note that in this case $k_{+}$ is not zero even for $\Lambda_b=0$ or 
$\alpha=\frac{1}{2}$.

The value of $k_+$ can be smaller or larger using the parameter 
$\alpha$ and the bulk cosmological constant $\Lambda_b$.  Smaller $k_+$ 
require longer interval length to explain the hierarchy between those scales, 
which results in lighter Kaluza-Klein (KK) modes of the graviton.   
As the KK modes must interact with the standard model particles through the 
gravitational interaction, the lighter KK mode has the longer lifetime. 
And sufficiently longer life time of the KK modes  
could have an effect on nucleosynthesis.  According to ref.\cite{chang}, 
the masses of the KK modes should be larger than about a few GeV, 
which corresponds to $k_+ b_0/2\lesssim 40$ to be consistent with the 
nucleosynthesis scenario.   
Therefore, smaller $k_+$ than $M$ cannot be consistent with the current 
cosmology.  

On the other hand, a larger $k_+$ corresponds to a larger curvature 
and then it would 
locate our theory out of perturbative regime.  Thus, it isn't desirable.  
We will show, however, in Sec.~VI that if quadratic curvature terms have the 
Gauss-Bonnet ratio, at least the quadratic corrections do not affect 
linearized 4-dimensional Einstein gravity or non-relativistic Newtonian 
gravity regardless of the curvature's magnitude.   

For $k_-b_0\simeq 74$, the -- solution has the same behavior.
The case of $\Lambda_b/M^5=12\alpha-6$ and $\alpha<\frac{1}{4}$ 
corresponds to $k_-=M\approx M_P$ and  
$\Lambda_1^{+}=-\Lambda_2^{-}=+6M^4|1-4\alpha|$.
Then we have only to set $b_0/2=37/M$ 
to explain the hierarchy between two scales too.
Of course, we also have the freedom to make the $k_-$ smaller or larger 
than $M$ depending on the parameters $\alpha$ and  
$\Lambda_b$.

The small warp factor \cite{rs} makes it possible to generate
a TeV scale mass from the fundamental parameter of
$O(M)$. But these TeV scale masses also appear in the other mass
parameters of the effective operators. In particular, the operators
leading to proton decay are also parametrized by a TeV scale
mass. Therefore, one has to suppress sufficiently the low
dimensional proton decay operators such that it is sufficiently 
long-lived ($\tau_p>10^{32}$~years), allowing operators with $D>14$
only.

In non-Gauss-Bonnet cases satisfying the condition
$16\alpha+5\beta+4\gamma=0$, the RS type solution is still valid 
except for the substitution $4\alpha\longrightarrow 8\alpha+\beta$
in Eq.~(21) because the RS metric can be 
redefined to be conformally flat.

\section{inflationary solutions}

For inflationary solutions we impose an ansatz, 
\begin{equation}
n=f(y), \quad a=g(\tau)f(y), \quad b=b_0,
\end{equation}
where $b_0$ is a constant.
Now adding the $(00)$ and $(ii)$ equations in Eq.~(\ref{Eeqn}),
we obtain 
\begin{equation}
-2\left(\frac{\dot g}{g}\right)^{\dot{\vphantom{g}}}
\left[1-\frac{4\alpha}{M^2b_{0}^2}\frac{f''}{f}\right]=0.
\end{equation}
Since $f''$ necessarily gives rise to a delta function 
we should take $(\dot g/g)\dot{\vphantom{g}}=0$.
So we define $(\dot g/g)\equiv H_0={\rm constant}$.
Then the $(55)$ equation gives
\begin{equation}
\left[\left(\frac{H_0}{f}\right)^2
-\left(\frac{f'}{f}\right)^2\frac{1}{b_{0}\,^2}\right]
+\frac{2\alpha}{M^2} \left[\left(\frac{H_0}{f}\right)^2
-\left(\frac{f'}{f}\right)^2\frac{1}{b_{0}\,^2}\right]^2
=\frac{\Lambda_b}{6M^3}~~~.
\end{equation}
After little algebra, we obtain 
\begin{equation}
\left(\frac{f'}{b_0}\right)^2 = H_0^2 + k_{\pm}^2 f^2,
\end{equation}
where the $k_{\pm}^2$ is defined in Eq.~(\ref{blcos}).   
For the `$-$' case with $\alpha=0$, we arrive at the solution
given in Ref.~\cite{kimkim}, $k_-^2=-\Lambda_b/6M^3$.    
{\it Note that $k_{+}^2$ contains the cases of 
both positive and negative cosmological constants in the bulk}  
and the $k_\pm^2$ can take both positive and negative signs. 
Inflationary solutions were obtained for a flat bulk geometry  
and for an AdS bulk geometry \cite{kal,nihei,lukas}, 
which can solve the hierarchy problem in the static limit.
In our case, as one can see below, inflationary 
solutions exist also
for a positive bulk cosmological constant.  

For $k_{\pm}^2>0$, the solution consistent with the orbifold
symmetry is
\begin{equation}
f = \frac{H_0}{k_{\pm}}\sinh(-k_{\pm}b_0|y|+c_0).
\end{equation}
The (00) or (ii) equations of Eq.~(\ref{Eeqn}) 
just give a boundary condition for the solution.  
Using the relation given in Eq.~(\ref{disconti}),
one can find that the imposed conditions determine the extra 
dimension scale $b_0$ and the integration constant $c_0$ as
follows,
\begin{eqnarray} 
k_{1,\mp}&=&\mp k_{\pm}\coth(c_0),\nonumber \\ 
k_{2,\pm}&=&\pm k_{\pm}\coth(-\frac{1}{2}kb_0+c_0)
\end{eqnarray}
where the $k_i$ are defined in Eq.~(\ref{brcos}).   
Here, the solutions are valid only for 
$k_{\pm}<|k_{1,\pm}|<|k_{2,\pm}|$ 
in case $c_0>\frac{1}{4}kb_0$ 
and $k_{\pm}<|k_{2,\pm}|<|k_{1,\pm}|$ 
in case $c_0<\frac{1}{4}kb_0$. One 
can also check easily that the $k_i$ tends to 
those of Ref.~{\cite{kimkim}} in the limit of $\alpha\rightarrow 0$ 
in the lower case (i.e. the -- solution).  
In general, inflation occurs if parameters $\alpha,\Lambda_b,\Lambda_1$,
and $\Lambda_2$ do not satisfy the two relations implied by
Eq.~(\ref{finetun}). From the above relations the 
$k_{1,\mp}(k_{2,\pm})$ diverges 
as $k_{2,\pm}(k_{1,\mp})\rightarrow \mp k_{\pm}{\rm coth}(\frac{1}{2}kb_0)$.  
 
Then the metric is
\begin{eqnarray}
\label{separable}
ds^2 &=& \left(\frac{H_0}{k_\pm}\right)^2\sinh^2(-k_\pm b_0|y|+c_0)
\left[-d\tau^2+e^{2H_0\tau}\delta_{ij}dx^idx^j\right]
+b_0^2dy^2.
\end{eqnarray}
To obtain the RS static solution with the warp factor in the 
visible brane, we should take 
$H_0\rightarrow 0$ and $c_0\rightarrow +\infty$ while keeping 
the ratio $(H_0e^{c_0})/(2k_{\pm})\rightarrow 1$ fixed.  
Then we obtain the fine tuning condition 
$k_{1,\mp}=-k_{2,\pm}=\mp k_{\pm}$ from Eq.~(\ref{brcos}), 
which is the same result as Eq.~(\ref{finetun}).  
Here one can see the possibility of the warp factored brane 
with the positive cosmological constant again.

After 4-dimensional coordinate transformation at a given $y$ to make the 
4-dimensional metric be in the form 
$ds_{4}^2=-dt^2+e^{2H(y)t}\delta_{ij}dx^idx^j$\cite{kimkim}, we get 
the hubble parameter expressed in
terms of the cosmological constant and the energy density,
$H_{\rm vis,\pm}=\sqrt{ (k_{\rm vis,\pm})^2
-k_{\pm}^2}$.  Here $k_{\rm vis,\pm}^2=k^2_\pm$ for the static
solutions and the two parameters corresponding to the + and -- solutions 
at the visible brane, $k_{\rm vis,\pm}$, are given by
\begin{equation} 
k_{\rm vis,\pm}={\left(\Lambda_2^\pm+\rho_{\rm vis}\right)\over
6M^3\sqrt{1+({4\alpha\Lambda_b}/{3M^5})}}
\end{equation}
where $\Lambda_2^\pm \gg\rho_{\rm vis}$.  
Thus the Hubble parameter at B2 is given by \cite{kkl}
\begin{equation} \label{friedman}
H_{\rm vis,\pm}^2=\frac{\rho_{\rm vis}(\rho_{\rm vis}+2\Lambda_2^\pm)
}{36M^6(1+4\alpha\Lambda_b/3M^5)}
=\frac{\pm \rho_{\rm vis}}{3M_{Pl}^2\sqrt{1+4\alpha\Lambda_b/3M^5}}
\left[1+\frac{\rho_{\rm vis}}{2\Lambda_2^\pm}\right]~~.
\end{equation}
The second equation above is derived with the use of the Eq.~(\ref{finetun}) 
and Eq.~(\ref{plank}).  
With $\rho_{\rm vis}=0$, we obtain the previous 
static solution. But with $\Lambda^-_2=\Lambda_b<0$
where the original RS solution sits, there exists a
possibility that $\rho_{\rm vis}(2\Lambda_2^-+\rho_{\rm vis})<0$ 
at a sufficiently low temperature, and hence it is
difficult to obtain a real Hubble parameter \cite{hubble,Lykken}. 
But with a positive
$\Lambda_2^+$, there does not exist such a problem.
This is possible for our + solution for $\alpha>0$. 
{\it Therefore, with the + solution
we can obtain a plausible Friedmann-Robertson-Walker
universe after inflation ends.}

In Eq.~(\ref{friedman}), the $\rho_{\rm vis}^2$ term gives a correction to  
the conventional Friedmann equation but in the limit 
$\Lambda_2^+\rightarrow \infty$ and $\alpha\Lambda_b/M^5\rightarrow 0$ 
we recover the standard 4-D general relativistic 
result.  The modified Friedmann equation leads to the modified inflation 
condition, 
\begin{equation}
\frac{\ddot{a}}{a}=\frac{-1}{3M_{Pl}^2\sqrt{1+4\alpha\Lambda_b/3M^5}}
\bigg(\frac{\rho_{\rm vis}}{2}(1+\frac{2\rho_{\rm vis}}{\Lambda_2^+})
+\frac{3p_{\rm vis}}{2}(1+\frac{\rho_{\rm vis}}{\Lambda_2^+})\bigg)>0
\end{equation}
or 
\begin{equation}\label{infcond}
p_{\rm vis}<-{\rho_{\rm vis}\over 3}
\left[{\Lambda_2^++2\rho_{\rm vis}\over \Lambda_2^++\rho_{\rm vis}}\right]
\end{equation}
where the $\rho_{\rm vis}$ and $p_{\rm vis}$ satisfy the fluid equation, 
$\dot{\rho}_{\rm vis}+3H(\rho_{\rm vis}+p_{\rm vis})=0$.  
 
Now let us consider the case that the only matter 
in the 4-D universe is a self 
interacting scalar field, inflaton $\phi$.  
Then the $\rho_{\rm vis}$ and $p_{\rm vis}$ are given by
 $\rho_{\rm vis}=\frac{1}{2}\dot{\phi}^2+V(\phi)$ and 
$p_{\rm vis}=\frac{1}{2}\dot{\phi}^2-V(\phi)$, 
respectively, and the Eq.~(\ref{infcond}) becomes~\cite{mwbh}
\begin{equation}
\dot{\phi}^2-V(\phi)+\left[\frac{\dot{\phi}^2+2V(\phi)}{8\Lambda_2^+}
(5\dot{\phi}^2-2V(\phi))\right]<0\,,
\end{equation}
which reduces to $\dot{\phi}^2<V$ when $\dot{\phi}^2+2V\ll\Lambda_{2}^+$.
Assuming that the inflaton field rolls down to a true vacuum 
very slowly, the energy 
density is dominated by the potential $V$ and the inflaton field evolution is 
strongly damped, which implies
\begin{eqnarray} \label{slowroll1}
H^2 &\simeq& \frac{V}{3M_{Pl}^2\sqrt{1+4\alpha\Lambda_b/3M^5}}
\left[1+{V\over2\Lambda_2^+} \right]\,\\ \label{slowroll2}
\dot\phi &\simeq&  -{V'\over 3H}\,,
\end{eqnarray}
where we use `$\simeq$' to denote equality within the slow-roll 
approximation~\cite{kolb}.  Our 
brane physics modifies also the e-folding number as follows;
\begin{equation}
N=\int_{t_{\rm i}}^{t_{\rm f}}dt H
\simeq \frac{-1}{M_{Pl}^2\sqrt{1+4\alpha\Lambda_b/3M^5}}
\int_{\phi_{\rm i}}^{\phi_{\rm f}}d\phi {V\over V'} 
\left[ 1+{V \over 2\Lambda_2^+}\right]\,. 
\end{equation}
For a large $\Lambda_2^+$ and small $\alpha\Lambda_b/M^5$ 
we obtain the standard e-folding formula again. 

Next, let us consider the corrections to the scalar and tensor 
density perturbations.  
The scalar density perturbation can be related to the curvature 
perturbation $\zeta$ on uniform density hypersurfaces 
when modes re-enter the Hubble radius  
during the matter dominated era~\cite{kolb,mwbh},
\begin{eqnarray}
A_{\sc s}^2 &=& \frac{4}{25}\langle \zeta^2 \rangle \nonumber \\
\zeta &=& {H\delta\phi\over\dot\phi}
\end{eqnarray}
where the scalar field fluctuation 
at Hubble crossing ($k=aH$) are given by 
$\langle\delta\phi^2\rangle\simeq\left({H/2\pi}\right)^2$.
Thus, using the slow-roll conditions Eq.~(\ref{slowroll1}) 
and Eq.~(\ref{slowroll2}), the amplitude of scalar 
perturbations becomes 
\begin{equation}
A_{\sc s}^2\simeq \frac{1}{75\pi^2}
\bigg(\frac{1}{M_{Pl}^2\sqrt{1+4\alpha\Lambda_b/3M^5}}\bigg)^3
{V^3\over V^{\prime2}}\left.\left[1+\frac{V}{2\Lambda_2^+}\right]^3
 \right|_{k=aH}\,.
\end{equation} 
So the amplitude of scalar perturbations is 
increased relative to the standard result. Of course, we recover
the standard one for a large $\Lambda_2^+$ and small $\alpha\Lambda_b/M^5$.

The amplitude of tensor (gravitational wave) perturbation
at Hubble crossing is given by~\cite{kolb,mwbh}
\begin{equation}
A_{\sc t}^2={1\over 50\pi^2}\left.\left(
{H\over M_{Pl}}\right)^2\right|_{k=aH}\,
\end{equation}
In the slow-roll approximation, this yields
\begin{equation}
A_{\sc t}^2\simeq \frac{1}{150\pi^2}
\bigg(\frac{1}{M_{Pl}^4\sqrt{1+4\alpha\Lambda_b/3M^5}}\bigg)
V\left.\left[1+\frac{V}{2\Lambda_2^+}\right]\right|_{k=aH}\,
\end{equation}
which is increased by brane effects, but with a smaller 
factor than in the case the scalar perturbation. This tends to the 
standard form as $\Lambda_2^+\rightarrow 
\infty$ and $\alpha\Lambda_b/M^5 \rightarrow 0$.

In the previous section, we discussed the phenomenologically
favored values of $\Lambda_b/M^5$, $\Lambda_2^+$ and $\alpha$.
The result was $k_+=M\approx M_P$, $\Lambda_b/M^5=12\alpha-6$, 
$\Lambda_2^+=6M^4|4\alpha-1|$ and $\alpha>\frac{1}{4}$.  
If we take $\Lambda_b=0$ or $\alpha=\frac{1}{2}$, 
we could recover the existing results in cosmology.

\section{other solutions}

If we take a non-separable ansatz for the metric tensor,
\begin{equation}
n(\tau,y)=a(\tau,y)=\frac{1}{\tau f(y)+g_0}, \quad
b(\tau,y)=k_{\pm} b_0\tau a(\tau,y),
\end{equation}
the solution is
\begin{equation} \label{nonsep}
ds^2 = \frac{-d\tau^2+\delta_{ij}dx^idx^j+(k_{\pm} b_0\tau)^2dy^2}%
{\left[k_{\pm} \tau\sinh(k_{\pm}b_0|y|+c_0)+g_0\right]^2},
\end{equation}
where $b_0$ and $c_0$ are constants and determined by
the boundary wall's conditions.  
\begin{eqnarray}
c_0 &=& \cosh^{-1}\left(\frac{\mp k_{1,\mp}}{k_{\pm}}\right), \nonumber\\
k_{\pm}b_0 &=& 2\left[\cosh^{-1}\left(\frac{\pm k_{2,\pm}}{k_{\pm}}\right)
              -\cosh^{-1}\left(\frac{\mp k_{1,\mp}}{k_{\pm}}\right)\right].
\end{eqnarray}
Actually this has the same form as in Ref.~\cite{kimkim}
except that the cosmological constants are, as before, 
given by Eq.~(\ref{blcos}) and Eq.~(\ref{brcos}).
The constant $g_0$ remains as a free parameter and its physical role is 
discussed in \cite{kimkim}.  Setting $g_0=0$, we get 
the solution given in Ref~\cite{nihei} and $b_0$ becomes 
independent of $\tau$.  

If $16\alpha+5\beta
+4\gamma=0$ is satisfied but the Gauss-Bonnet conditions are not,
the inflationary solutions in given Eq.~(\ref{separable}) 
(with a separable metric) and Eq.~(\ref{nonsep}) (with a 
nonseparable metric) are still valid except the substitution $4\alpha
\longrightarrow 8\alpha+\beta$ in our solutions Eq.~(\ref{blcos}), etc., 
even though there 
exist higher time derivatives in the equations of motion.  

Taking a different ansatz, $n(\tau,y)=a(\tau,y)=b(\tau,y)$, 
which is conformally flat, the solution is given by 
\begin{equation}
ds^2 = \frac{-d\tau^2+\delta_{ij}dx^idx^j+dy^2}%
{\left[-(k_{1,\mp}^2-k_{\pm}^2)^{1/2}\tau+k_{1,\mp}|y|+c_0\right]^2},
\end{equation}
where $c_0$ is a constant.  This metric describes inflation 
in both the spatial dimensions and the extra dimension.
The $k$'s are given by Eq.~(\ref{blcos})and Eq.~(\ref{brcos}) 
as before. 
For non-Gauss-Bonnet case with $16\alpha+5\beta+4\gamma=0$, 
the solution is still valid also except for the substitution 
$4\alpha\longrightarrow 8\alpha+\beta$ because it is conformally flat.  

\section{metric perturbation near the RS background geometry}

It is also of interest to study the gravitational interaction with
the RS background. In fact, Randall and Sundrum
demonstrated that the Newton's force law does not imply only 
four non-compact dimensions in the presence of a non-factorizable 
background geometry~\cite{rs2,longi}.  The example they studied is 
the case of a {\it single} 3-brane embedded in non-compact five dimension.  
In this section, let us reconsider the case with the Gauss-Bonnet 
interaction.  

The graviton is a linearized tensor fluctuation near the background geometry,  
\begin{equation} \label{h}
g_{\mu \nu} = e^{-2 k_{\pm}|y|} \eta_{\mu \nu} + h_{\mu \nu}(x,y).
\end{equation}
where the $x$ indicates the coordinate for the 4-dimensional space 
embedded in the 5-dimensional bulk.
Since we are interested in the 4-dimensional graviton only, 
which is the longitudinal component of the metric 
fluctuation, we set $h_{\mu 5}=h_{5 \mu}=h_{55}=0$.  
Inserting  Eq.~(\ref{h}) into Eq.~(\ref{eom}) and 
taking only the linear terms in $h_{\mu \nu}$, we obtain  
\begin{eqnarray}\label{lineareins}
\underline{G_{\mu \nu}}+\underline{X_{\mu \nu}}&=&
\bigg[-\frac{1}{2}\left(1-\frac{4\alpha k_{\pm}^2}{M^2}\right)\partial_y^2
+\frac{8\alpha k_{\pm}^2}{M^2}\delta(y) sgn(y)\partial_y \nonumber \\
&&-\frac{\Box_4}{2}e^{2 k_{\pm}|y|}\left(1-\frac{4\alpha k_{\pm}^2}{M^2}
+\frac{8\alpha k_{\pm}}{M^2}\delta(y)\right) \nonumber \\
&&-8k_{\pm}\delta(y) \left(1-\frac{6\alpha k_{\pm}^2}{M^2} \right)
+4k_{\pm}^2\left(2-\frac{5\alpha k_{\pm}^2}{M^2}\right)\bigg]h_{\mu\nu}(x,y)
\end{eqnarray}
and 
\begin{eqnarray}\label{linearenergy}
\underline{T_{\mu \nu}}&=&
-\frac{1}{M^3}\left[\Lambda_b+\Lambda_1^{\mp}\delta(y)\right]
h_{\mu \nu}(x,y)\nonumber \\
&=&\left[6k_{\pm}^2\left(1-\frac{2\alpha k_{\pm}^2}{M^2}\right)
-6k_{\pm}\delta(y)\left(1-\frac{4\alpha k_{\pm}^2}{M^2}\right)
\right]h_{\mu\nu}(x,y)
\end{eqnarray}
where the underlined quantities denote the linear part in 
$h_{\mu \nu}$ in the full expressions
and $\Box_4$ is $\eta^{\mu\nu}\partial_{\mu}\partial_{\nu}$. 
Here we set $b_0=1$ for simplicity.  
Eq.~(\ref{linearenergy}) is obtained by the use of Eq.~(\ref{blcos}) 
and Eq.~(\ref{finetun}). Here we choose the traceless transverse
gauge conditions, 
$\partial^{\mu}h_{\mu \nu} = h^{\mu}_{\mu} = 0$ \cite{rs2,longi}.  
Under this gauge condition all components of $h_{\mu \nu}$ 
satisfy the same equation of motion, and hence we will omit the 
$\mu \nu$ indices below. 
Here we note again that in the Gauss-Bonnet case 
the unwanted higher derivative terms  
disappear in the linear approximation as in the background case.

To perform a Kaluza-Klein reduction down to 4-dimension and 
get an understanding of all modes that appear in 
the assumed 4D effective theory, we seperate the variables; 
$h(x,y)=\psi(y)e^{i p \cdot x}$, where the $p^{\mu}$ 
is a 4-dimensional momentum. Since the 4-dimensional
mass $m^2$ of the KK excitation is $p^2=-m^2$,   
Eq.~(\ref{lineareins}) = Eq.~(\ref{linearenergy}) leads to 
\begin{eqnarray}\label{linearized}
\bigg[
&-&{1 \over 2}\left(1-\frac{4\alpha k_{\pm}^2}{M^2}\right) \partial_y^2
+2k_{\pm}^2\left(1-\frac{4\alpha k_{\pm}^2}{M^2}\right) \nonumber \\
&-&2k_{\pm} \delta(y)\left(
-\frac{4\alpha k_{\pm}}{M^2}sgn(y)\partial_y
+1-\frac{12\alpha k_{\pm}^2}{M^2}\right)\bigg]\psi(y) 
\nonumber \\
&&={m^2 \over 2} e^{2 k_{\pm}|y|}\bigg[1-\frac{4\alpha k_{\pm}^2}{M^2}
+\frac{8\alpha k_{\pm}}{M^2}\delta(y)\bigg]\psi(y).
\end{eqnarray}
Note that the above equation remains the same regardless 
of the sign of the brane's cosmological constant.

In the bulk, we can easily check that the Gauss-Bonnet interaction does not 
modify the equation of motion because all terms have the exactly same common 
factor $(1-4\alpha k_{\pm}^2/M^2)$ neglecting the Dirac delta functions.  
Therefore, we obtain {\it the same eigenfunctions and eigenvalues as in 
the RS's solutions}~\cite{rs2} except for the definition of $k_{\pm}$.  
On the other hand, the Dirac delta functions gives the boundary 
condition, so the Gauss-Bonnet interaction modifies 
only the boundary condition in the order of magnitude of 
$\alpha k_{\pm}^2/M^2$. Thus, one can imagine that there exists 
a possibility that the massless mode has not only an exponentially 
decaying component but also an exponentially growing one at order 
$\alpha k_{\pm}^2/M^2$. Note that the coefficient 
of the growing mode is exactly zero in the absence of 
the higher curvature terms in the action~\cite{rs2}.  
We have found, however, that the exponentially growing mode does not 
appear even in the presence of the Gauss-Bonnet interaction.  

To follow the RS process, let us make change of variables; 
$z \equiv sgn(y) \left( e^{k_{\pm} |y|} -1 \right)/k_{\pm}$, $\hat{\psi}(z) 
\equiv\psi(y)e^{k_{\pm}|y|/2}$ and $\hat{h}(x,z)
\equiv h(x,y)e^{k_{\pm}|y|/2}$. Then, Eq.~(\ref{linearized}) reads 
\begin{eqnarray} \label{finaleq}
\bigg[-{1 \over 2}\partial_z^2 +{15 k_{\pm}^2 \over 8 (k_{\pm}|z|+1)^2}
&+&{k_{\pm}\over 2}\delta(z)\left(\frac{2B}{Ak_{\pm}}sgn(z)\partial_z
-\frac{3C}{A}\right)\bigg] \hat{\psi}(z) \nonumber \\ 
&=& \frac{m^2}{2}\left[1+\frac{B}{Ak_{\pm}}\delta(z)\right] \hat{\psi}(z),
\end{eqnarray}
where 
$A=1-4\alpha k_{\pm}^2/M^2$, $B=8\alpha k_{\pm}^2/M^2$  
and $C=1-12\alpha k_{\pm}^2/M^2$.  Note that $A=B+C$.    

For $m^2=0$, the eigenfunctions in the bulk satisfying the
orbifold symmetry are 
$(k_{\pm}|z|+1)^{-3/2}/k_{\pm} (=\exp(-\frac{3}{2}
k_{\pm}|y|)/k_{\pm})$ and 
$(k_{\pm}|z|+1)^{5/2}/k_{\pm} 
(=\exp(\frac{5}{2}k_{\pm}|y|)/k_{\pm})$.
Therefore, the solution is a linear combination of them
\begin{equation} \label{massless}
a\frac{(k_{\pm}|z|+1)^{-3/2}}{k_{\pm}}+b\frac{(k_{\pm}|z|
+1)^{5/2}}{k_{\pm}}.
\end{equation}
To satisfy the boundary condition at $z=0$, let us 
insert Eq.~(\ref{massless}) 
into Eq.~(\ref{finaleq}) and assemble the coefficients of 
the Dirac delta functions. Then, we obtain
\begin{equation}
a(-A+B+C)+b(A-B+C)=0
\end{equation}
where $A=B+C$. Thus, $b=0$, i.e.  
the massless graviton is confined on the brane 
and the Newton's force law on the brane holds good {\it even under the 
non-compact extra dimension}. 
Particularly, we note that the result is not changed even though the brane's 
cosmological constant is negative.  

As $h_{\mu\nu}(x,y)\propto e^{-\frac{k_{\pm}}{2}|y|}\hat{\psi}(z)e^{ipx}
\propto e^{-2k_{\pm}|y|}e^{ipx}$, 
the fluctuation near the background metric can be written down as 
\begin{equation}\label{fl}
g_{\mu\nu}^0=e^{-2k_{\pm}|y|}(\eta_{\mu\nu}+\epsilon_{\mu\nu}e^{ipx}),
\end{equation} 
where the superscript $0$ denotes massless fluctuation and 
the $\epsilon_{\mu\nu}$ is a polarization tensor 
of the graviton wave function and from which we can see that the massless 
mode fluctuates only in the longitudinal direction to the brane.  
Besides, from Eq.~(\ref{linearized}) we can get 
the 4-dimensional linearized Einstein equation in the Minkowski space,  
\begin{equation}\label{4dlee}
-\frac{\Box_4}{2}\epsilon_{\mu\nu}e^{ipx}=0,
\end{equation}
which is of course true also after 4-dimensional coordinate transformation 
at a given $y$ to make the 4-dimensional background metric be in the form 
$dS_4^2=\eta_{\mu\nu}dx^{\mu}dx^{\nu}$.  
Note that {\it the above result is not affected 
by the Gauss-Bonnet correction}.   

We usually worry about the instability of anti-de Sitter space 
due to excitations of ghost particles \cite{deser}. 
In our case, we still have such a problem 
since the sign of the kinetic term for the $k_+$ background is 
opposite to that of the case without the Gauss-Bonnet term, viz.
\begin{equation}
-\frac{1}{2}\bigg(1-\frac{4\alpha k^2_{\pm}}{M^2}\bigg)
(e^{2k_\pm |y|}\Box_4+\partial^2_y)h_{\mu\nu}
=\pm\frac{1}{2}\sqrt{1+\frac{4\alpha\Lambda_b}{3M^5}}
(e^{2k_\pm |y|}\Box_4+\partial^2_y)h_{\mu\nu}.
\end{equation}
However, the equation of motion itself describes the same behavior of gravity 
localization on the hidden sector brane (B1) because the brane cosmological 
constant contributing to energy momentum tensor changes its sign as well.  
Thus, we have no ghost problem as far as the brane cosmological constant is 
concerned as energy density.  But we cannot regard the brane with 
negative cosmological constant at $z=0$ as our universe due to the later 
cosmological problem that was discussed in Sec.~IV.

For $m^2>0$, the solutions for the above equation of motion 
in the bulk are  
\begin{equation}\label{massivesol}
a(|z|+1/k_{\pm})^{1/2}J_2(m(|z|+1/k_{\pm}))
+b(|z|+1/k_{\pm})^{1/2}Y_2(m(|z|+1/k_{\pm}))
\end{equation}
which is the same solution as in the RS case except for the 
definition of $k_{\pm}$. The imposed boundary condition at $z=0$ 
fixes the ratio of $a$ and $b$,
\begin{equation}
\frac{a}{b}=\frac{4k_{\pm}^2}{\pi m^2}\times
\frac{1-4\alpha k_{\pm}^2/M^2+2\alpha m^2/M^2}
{1-12\alpha k_{\pm}^2/M^2+\alpha m^2/M^2}.
\end{equation}  
In this case, the Gauss-Bonnet interaction modifies
Newton's non-relativistic gravitational potential {\it through 
the KK states} as follows,  
\begin{eqnarray}
V\sim G_N\frac{m_1m_2}{r}&+&\pi\int_{0}^{\infty}
\frac{dm}{k_{\pm}}\frac{m}{k_{\pm}}G_N\frac{m_1m_2e^{-mr}}{r}
\times \bigg[\frac{1-12\alpha k_{\pm}^2/M^2+\alpha m^2/M^2}
{1-4\alpha k_{\pm}^2/M^2+2\alpha m^2/M^2}\bigg]^2 \nonumber \\
\sim &G&_N\frac{m_1m_2}{r}\bigg[1+\frac{\pi}{(k_{\pm}r)^2}
\times \left(\frac{1-12\alpha k_{\pm}^2/M^2}{1-4\alpha k_{\pm}^2/M^2}
\right)^2 \bigg].
\end{eqnarray} 
The above result is obtained through the RS
technique given in Ref.~\cite{rs2}.
Of course, the potential given above is not ruled out yet~\cite{Lykken}.  

For the case of two branes and bulk with $S^1/Z_2$ symmetry, 
the Eq.~(\ref{linearized}) is modified into 
\begin{eqnarray}\label{2lin}
\bigg[
&-&{1 \over 2}\left(1-\frac{4\alpha k_{\pm}^2}{M^2}\right) \partial_y^2
+2k_{\pm}^2\left(1-\frac{4\alpha k_{\pm}^2}{M^2}\right) \nonumber \\
&-&2k_{\pm} \delta(y)\left(
-\frac{4\alpha k_{\pm}}{M^2}sgn(y)\partial_y
+1-\frac{12\alpha k_{\pm}^2}{M^2}\right)\nonumber \\
&+&2k_{\pm} \delta(y-\frac{1}{2})\left(
-\frac{4\alpha k_{\pm}}{M^2}sgn(y)\partial_y
+1-\frac{12\alpha k_{\pm}^2}{M^2}\right)\bigg]\psi(y) 
\nonumber \\
&&={m^2 \over 2} e^{2 k_{\pm}|y|}\bigg[1-\frac{4\alpha k_{\pm}^2}{M^2}
+\frac{8\alpha k_{\pm}}{M^2}\delta(y)
-\frac{8\alpha k_{\pm}}{M^2}\delta(y-\frac{1}{2})\bigg]\psi(y)
\end{eqnarray}
where the $sgn(y)\equiv |y|'=2(\theta(y)-\theta(y-\frac{1}{2}))-1$ 
and we can check the solution Eq.~(\ref{massless}) with $b=0$
satisfies the above equation 
(\ref{2lin}) regardless of the lengh scale by use of the relation 
$h_{\mu\nu}=e^{-\frac{k_{\pm}}{2}|y|}\hat{\psi}(z)$ and Eq~.(\ref{disconti}).  
Therefore, from Eq.(\ref{4dlee}) the non-relativistic Newtonian gravity 
could be restored at the visible sector brane for sufficiently 
small interval length $b_0<<r$. 

\section{conclusion}

We studied various static and inflationary solutions
in the Randall-Sundrum framework
with the Gauss-Bonnet term added to the standard Hilbert
action. It has been argued that in this RS framework the
Gauss-Bonnet term is the only acceptable curvature square
term. Then there exist additional coupling $\alpha$, the 
coefficient of the Gauss-Bonnet term. Depending on various 
values of $\alpha$, there exist static solutions and also
the inflationary solutions. In particular, there exist
solutions for a positive visible sector tension $\Lambda_2$
for $\alpha>0$, which makes it possible to transit to a standard
Big Bang cosmology after inflation.

\acknowledgments
We thank K. Choi, S. Moon and J. D. Park for useful discussions.
This work is supported in part by the BK21 program of Ministry 
of Education, and by the Korea Science and Engineering Foundation
(KOSEF 1999 G 0100).

\newpage 
\begin{figure}
\epsfxsize=160mm
\centerline{\epsfbox{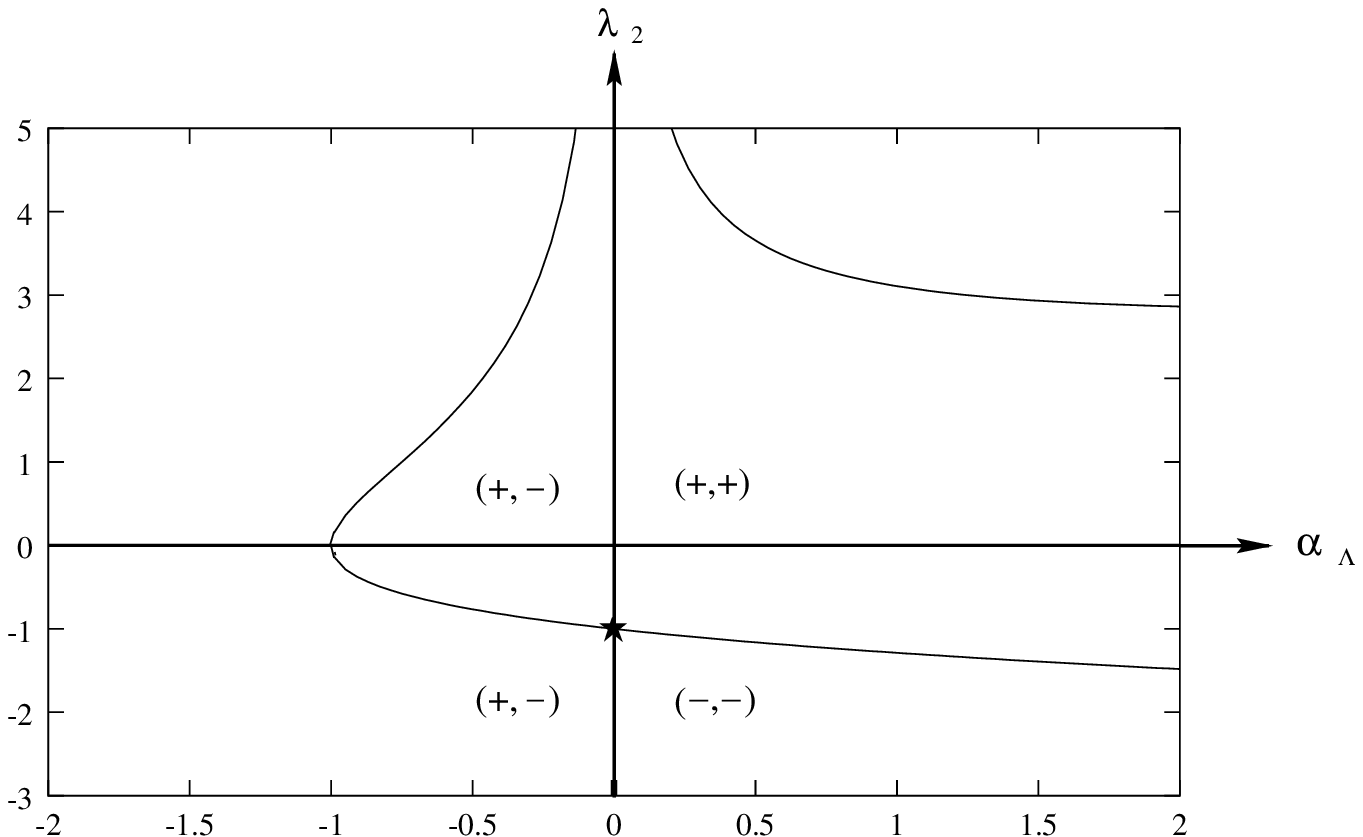}}
\end{figure}
\noindent Fig.~1. Possible solutions for $\lambda_2\equiv
\Lambda_2/\sqrt{6M^3|\Lambda_b|}$  as a function of 
$\alpha_{\Lambda}\equiv 4\alpha\Lambda_b/(3M^5)$. The star point
is the RS solution. The four quadrants have different sets of
signs of $\alpha$ and $\Lambda_b$, denoted as $({\rm sign\ of\ }
\alpha,\ {\rm sign\ of\ } \Lambda_b)$. 





\begin{references}

\bibitem{rs} L. Randall and R. Sundrum, Phys. Rev. Lett.
{\bf 83}, 3370 (1999) [hep-ph/9905221].

\bibitem{kal} N. Kaloper, Phys. Rev. {\bf D60}, 123506 (1999) [hep-th/9905210]. 

\bibitem{nihei} T. Nihei, Phys. Lett. {\bf B465}, 81 (1999) [hep-ph/9905487].

\bibitem{kimkim} H. B. Kim and H. D. Kim, hep-ph/9909053. 

\bibitem{kkl} J. E. Kim, B. Kyae and H. M. Lee, hep-ph/9912344

\bibitem{orb} L. Dixon, J. A. Harvey, C. Vafa, and E. Witten,
Nucl. Phys. {\bf B261}, 678 (1985); ibid.   
Nucl. Phys. {\bf B274}, 285 (1986).

\bibitem{wise} W. D. Goldberger and M. B. Wise, Phys. Rev. Lett. {\bf 83}, 4922 
(1999) [hep-ph/9907447]

\bibitem{inflation} Notations for $k$'s are borrowed from
Refs.~\cite{nihei,kimkim}.

\bibitem{ghost} B. Zwiebach, Phys. Lett. {\bf B156}, 315 (1985).

\bibitem{deser} D.G. Boulware and S. Deser, Phys. Rev. Lett. {\bf 55}, 2656 (1985).

\bibitem{shapiro} A.L. Maroto and I.L. Shapiro, Phys. Lett. {\bf B414}, 34 (1997).

\bibitem{chang} S. Chang and M. Yamaguchi, hep-ph/9909523

\bibitem{hubble} P. Binetruy, C. Deffayet and D. Langlois, hep-th/9905012;
C. Cs$\acute{\rm a}$ki, M Graesser, C. Kolda, and J. Terning, hep-ph/9906513;
J. M. Cline, C. Grojean and G. Servant, Phys. Rev. Lett. 83, 4245 (1999) 
[hep-ph/9906523];
C. Cs$\acute{\rm a}$ki, M. Graesser, L. Randall, and J. Terning,
Phys. Lett. {\bf B462}, 34 (1999) [hep-ph/9911406]. 

\bibitem{Lykken} J. Lykken and L. Randall, hep-th/9908076.

\bibitem{lukas} A. Lukas, B. A. Ovrut and D. Waldram,
Phys. Rev. {\bf D61}, 023506 (2000)[hep-th/9902071]. 

\bibitem{kolb} J. E. Lidsey, A. R. Liddle, E. W. Kolb, 
E. J. Copeland, T. Barreiro
and M. Abney, Rev. Mod. Phys. {\bf 69}, 373 (1997) [astro-ph/9508078].

\bibitem{mwbh} R. Maartens, D. Wands, B. A. Bassett and 
I. Heard, hep-ph/9912464.  

\bibitem{rs2} L. Randall and R. Sundrum, Phys. Rev. Lett.
{\bf 83}, 4690 (1999) [hep-th/9906064]; 
N. Arkani-Hamed, S. Dimopoulos, G. Dvali and N. Kaloper, hep-th/9907209;
C. Cs$\acute{\rm a}$ki, J. Erlich, T. J. Hollowood 
and Y. Shirman, hep-th/0001033; 
S. B. Giddings, E. Katz and L. Randall, hep-th/0002091.

\bibitem{longi} 
M. G. Ivanov and I. V. Volovich, hep-th/9912242; 
Y. S. Myung and G. Kang, hep-th/0001003; 
Y. S. Myung, G. Kang and H. W. Lee, hep-th/0001107.

\end{references}
\end{document}